\PassOptionsToPackage{pdfpagelabels=false}{hyperref}
\documentclass[fleqn,usenatbib]{mnras}
\usepackage{newtxtext,newtxmath}

\usepackage[T1]{fontenc}
\usepackage{ae,aecompl}

\usepackage{graphicx}	
\usepackage{amsmath}	
\usepackage{amssymb}	

\newcommand{\CO}{$^{12}$CO}
\newcommand{\CIt}{$\mathrm{^3P_2 - ^3P_1}$}
\newcommand{\CIo}{$\mathrm{^3P_1 - ^3P_0}$}

\defcitealias{FFtech}{Paper~I}
\defcitealias{FFredshift}{Paper~II}
\defcitealias{FFlineID}{Paper~III}
\defcitealias{FFncc}{Paper~IV}

\title[SPIRE FF IV: Neutral Carbon Detection]{The \textit{Herschel} SPIRE Fourier Transform Spectrometer Spectral Feature Finder IV. 
Neutral Carbon Detection in the SPIRE FTS Spectra\thanks{\textit{Herschel} was an ESA space observatory with science instruments provided by European-led Principal Investigator consortia and with important participation from NASA.}}

\author[J. P. Scott et al.]{Jeremy P. Scott,$^{1}$\thanks{E-mail: jeremy.scott@uleth.ca}
Locke D. Spencer,$^{1}$
Rosalind Hopwood,$^{2,3}$
Ivan Valtchanov,$^{2}$
\newauthor
and David A. Naylor.$^{1}$
\\
$^{1}$Department of Physics \& Astronomy, University of Lethbridge, AB, T1K 3M4, Canada\\
$^{2}$Telespazio Vega UK for ESA, European Space Astronomy Centre, Operations Department, 28691 Villanueva de la Ca\~nada, Spain\\
$^{3}$ Department of Physics, Imperial College London, Prince Consort Road, London SW7 2AZ, UK\\
}

\date{Accepted Jun.\ 2020. Received Jun.\ 2020; in original form Feb.\ 2020}

\pubyear{2020}

\begin{document}
\label{firstpage}
\pagerange{\pageref{firstpage}--\pageref{lastpage}}
\maketitle

\begin{abstract}
The SPIRE FTS Spectral Feature Finder (FF), developed within the \emph{Herschel} Spectral and Photometric Imaging Receiver (SPIRE) Fourier Transform Spectrometer (FTS) instrument team, is an automated spectral feature fitting routine that attempts to find significant features in SPIRE FTS spectra. The \CIo\ and \CIt\ neutral carbon fine structure lines are common features in carbon rich far-infrared astrophysical sources. These features can be difficult to detect using an automated feature detection routine due to their typically low amplitude and line blending. In this paper we describe and validate the FF sub-routine designed to detect the neutral carbon emission observed in SPIRE spectral data.
\end{abstract}




\section{Introduction}
The Herschel Space Observatory (\textit{Herschel}; \cite{Pilbratt10}) is an observatory class mission of the European Space Agency (ESA). Equipped with a 3.5\,m primary mirror, which is passively cooled to $ \sim $80\,K, \emph{Herschel} accessed an unobstructed view of the Far-Infrared (FIR) region of the electromagnetic spectrum with unprecedented angular resolution from orbit around the L2 Lagrangian point. Launched on May 14, 2009, \emph{Herschel} explored the FIR/submillimetre universe using a suite of three instruments including the Spectral and Photometric Imaging Receiver (SPIRE; \cite{Griffin10}), which consists of a three-band imaging photometer and an imaging Fourier Transform Spectrometer (FTS; \cite{spencer2015frequency}). The SPIRE FTS consists of two detector arrays, the SPIRE Long Wavelength (SLW) and SPIRE Short Wavelength (SSW) arrays, sensitive to 447-1018\,GHz and 944-1568\,GHz radiation, respectively \citep{spire_handbook}, providing broad, simultaneous spectral coverage. The spectrometer mirror mechanism scan range controls the spectral resolution of the instrument and operated in two modes for science observations, offering low / high resolution spectra with resultant frequency resolutions of $\Delta f =$ 24.98 / 1.2\,GHz, respectively.\@ The SPIRE photometer and spectrometer detectors were actively cooled by a closed cycle $^3$He / $^4$He system in order to achieve optimal detection sensitivity at an operating temperature of $\sim$300\,mK. With the best spatial resolution, broad spectral coverage, and high sensitivity at moderate spectral resolution for FIR observations, the \emph{Herschel} space mission targeted the science goals of studying the composition and dynamics of the interstellar medium, protostars, and galaxies, providing insight into star formation and the physical/chemical evolution of the Universe.

\begin{figure*}
\centering
\includegraphics[width=1.0\textwidth]{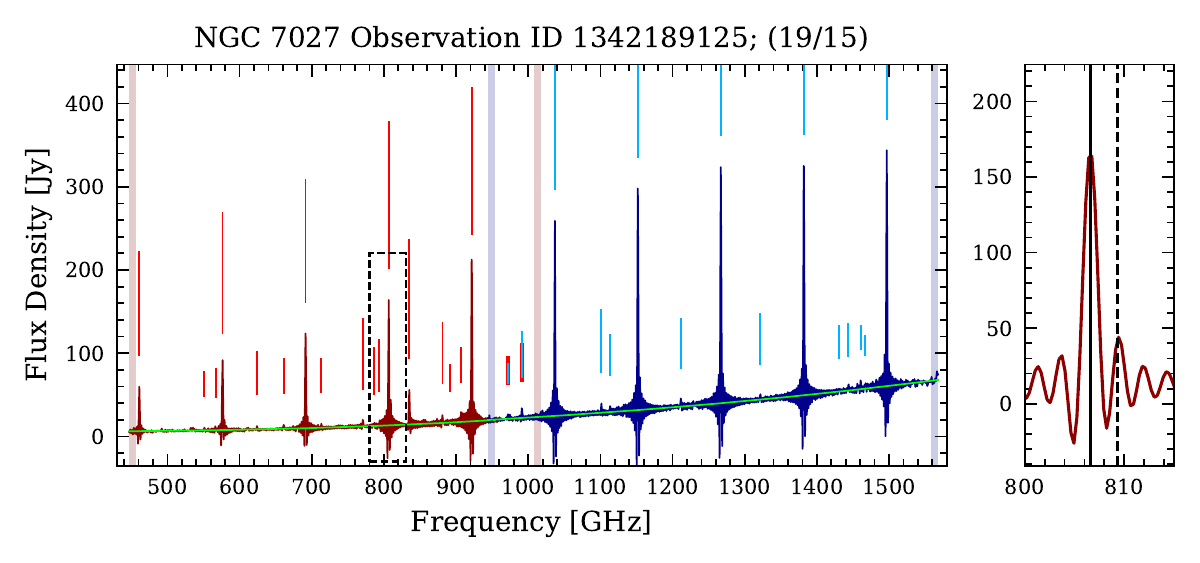}
\vspace{-18pt}
 \caption{A postcard generated by the FF for one NGC 7027 observation using standard HSA data after being processed by the FF with the NCC subroutine turned off. The red and blue spectra represent level-2 processed data from the SLW and SSW central detectors, respectively. The vertical bars indicate the position of detected features with the length of the bars indicating the relative signal to noise ratio of the associated features. The fitted continuum is shown with the green curve. To the right is a close-up of the region under investigation, with the detected \CO(J=7--6) line indicated by the solid line, and the expected position of the undetected [CI]\CIt\ line indicated by the dashed line.}
 \label{fig:NGC7027_FF}
\end{figure*}

The SPIRE instrument has produced a wealth of publicly available science-ready data, including spectra and hyperspectral images, accessible through the Herschel Science Archive (HSA). Analysis of this spectral data, however, can be time consuming due partially to the sinc-like instrument line shape (ILS) of the SPIRE spectrometer. To help with data mining and preliminary analysis, the SPIRE instrument team has developed the SPIRE Spectral Feature Finder \cite[FF; \citetalias{FFtech}:][]{FFtech} to automatically detect significant spectral features and determine line parameters within high resolution SPIRE FTS data (Fig.\,\ref{fig:NGC7027_FF}). The FF has produced a publicly available, searchable, catalogue\footnote{\url{https://www.cosmos.esa.int/web/herschel/spire-spectral-feature-catalogue}} as part of the HSA. A specific challenge faced by the FF routine is the task of finding emission lines that are in close spectral proximity to other prominent features. A particular example is the neutral carbon emission line ([CI]\CIt\ with rest frequency 809.34\,GHz) adjacent to the carbon monoxide emission features (\CO(J=7--6) with rest frequency 806.65\,GHz; Fig.\,\ref{fig:NGC7027_FF}). These two features are separated by 2.69\,GHz (2.24 resolution units) in the rest frame, superimposing the [CI]\CIt\ feature onto the first higher frequency positive side-lobe of the \CO(J=7--6) line. In addition, it is often the case that the [CI]\CIt\ feature is a fraction of the amplitude of the \CO(J=7--6) feature. As a result, it is difficult to distinguish the [CI]\CIt\ line from the \CO(J=7--6) side-lobe. Further difficulty is introduced by the asymmetric line shape observed in SPIRE data \citep{Naylor:16}, where the first higher frequency side-lobe often exhibits a positive flux offset with respect to the corresponding lower frequency side-lobe. 

Detection of the [CI]\CIo\ fine-structure line of neutral carbon, emitted at a rest frequency of 492.16\,GHz, can also be challenging to detect in SPIRE spectra using an automated routine. The difficulty arises from the typically low amplitude of this feature, in addition to its location in the noisy low frequency edge of the SLW band. 

The \CO\ molecule is of particular interest due to its high abundance in the interstellar medium \citep{herbst1995chemistry}. The rotational emission spectrum is especially bright in the SPIRE SLW and SSW bands, capturing ten rest frame emission lines from J=4--3 up to J=13--12. \CO\ is also useful as an H$_2$ tracer and helps map the distribution and velocity fields of cold gas in molecular clouds, revealing structures in evolving protostars such as protostellar disks and bipolar outflows (\cite{thi2001h2} and \cite{eisloffel2012sofia}). Neutral carbon is another species used to probe the interstellar medium around star forming regions. The ground state of neutral carbon (1s$^2$2s$^2$2p$^2$) is split into five fine-structure levels with $^3$P$_1$ and $^3$P$_2$ being only 2.04\,meV and 5.38\,meV above the ground state, respectively. These states can be populated by collisions in gas clouds and are appropriate tracers of photo-dissociation regions around young stellar objects \citep{tomassetti2014atomic}. 

The most prominent generator of neutral carbon is the photo-dissociation of other carbon containing molecules in circumstellar envelopes \citep{huggins1982photochemistry}. Neutral carbon tends to occur in low abundance with the exception of localized regions in circumstellar clouds \citep{huggins1982photochemistry}. This is because carbon that is cold enough to exist in its neutral states can easily bind to other atoms and form molecules. Additionally, the first ionized state of carbon is only $\sim$0.15\,eV greater than the dissociation energy of \CO. Neutral carbon is therefore restricted to a narrow energy range and explains why the [CI]\CIt\ line is often observed with a lower amplitudes than the adjacent \CO(J=7--6) line.

The Neutral Carbon Check (NCC) is a dedicated automated search algorithm used to help detect the neutral carbon [CI]\CIt\ and \CIo\ emission features in SPIRE FTS spectra as part of the FF \citepalias{FFtech}. This routine is outlined in \S\,\ref{sec:Method}. We then validate this routine using manual inspection and redundant feature detection in \S\,\ref{sec:Validation}. Results as they pertain to the SPIRE spectral feature catalogue are presented in \S\,\ref{sec:Results}. Concluding remarks are presented in \S\,\ref{sec:Conclusion}.

This paper is one of a series of four which discuss aspects of the FF: the main technical paper \citep[\citetalias{FFtech}:][]{FFtech}, a radial source velocity paper \citep[\citetalias{FFredshift}:][]{FFredshift}, a line identification paper \citep[\citetalias{FFlineID}:][]{FFlineID} which also presents FF results for the off-axis spectra within sparse observations, and this [CI] detection and deblending paper.

\section{Method}
\label{sec:Method}

A detailed outline of FF operations and performance is discussed in \citetalias{FFtech}. Here we simply highlight relevant validation results. The FF has demonstrated a 98\% detection rate of lines with a signal-to-noise ratio (SNR) $ \ge 10 $ using a large dataset of 20,000 simulated spectra containing the \CO\ rotational emission ladder. Additionally, radial velocity estimates meeting the FF threshold criteria have demonstrated a residual velocity of $ < $20\,km/s for $ \sim $90\% of estimates when compared to test case datasets consisting of real and simulated data \citep{FFredshift}. Detection of \CO\ and accurate radial velocity determination is crucial for successful application of the NCC.

The NCC routine initiates by obtaining a list of features and associated parameters from the FF catalogue for the source being examined. Then, all absorption features are removed. The new line list is then searched separately for features residing within 0.8\,GHz of the velocity adjusted frequency of both \CO(J=7--6) and [CI]\CIt. This range corresponds to a velocity error of $ \sim $300 km/s, yet is likely sufficiently constrained as to not confuse the expected features with emission from other species. If only one of these features is found, a 40\,GHz spectral window is isolated, centred on this feature. We do not discriminate based on which feature is found since the rare case exists where the [CI]\CIt\ feature is the more prominent detected feature with the \CO(J=7--6) feature being neglected. If both features are found prior to the NCC, no attempt is made to refine feature detection.

\begin{figure}
\centering
\vspace{-6pt}
\includegraphics[width=1.0\linewidth]{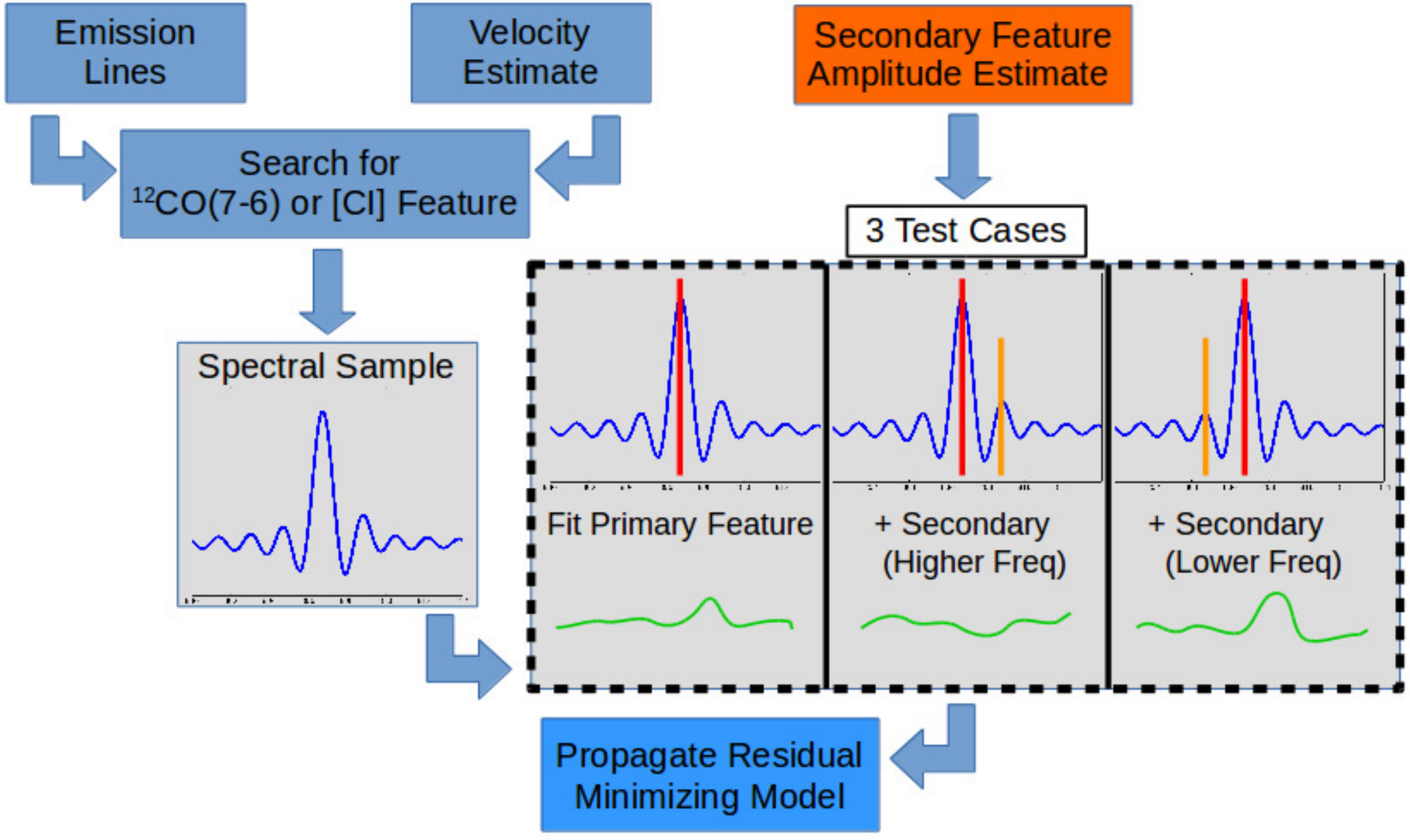}
\vspace{-15pt}
\caption{A flowchart of the NCC method. FF emission lines are searched for features within 0.8\,GHz of the expected velocity corrected frequency of \CO(J=7--6) or [CI]\CIt. If one feature is found, a spectral sample is extracted about the feature and subjected to 3 test cases. The model that minimizes the residual is considered the best representation of the spectral data.\vspace{-12pt}}
\label{fig:NeutralC_FlowChart}
\end{figure}

After the 40\,GHz spectral sample is isolated, three models are tested:
\begin{enumerate}
\item One primary feature (the initially detected \CO(J=7--6) or [CI]\CIt\ line), the continuum as determined by earlier steps in the FF, and all other features detected by the FF residing within this spectral sample. 
\item Model (i) with an additional secondary feature placed at a greater frequency relative to the primary feature.
\item Model (i) with an additional secondary feature placed at a lower frequency relative to the primary feature.
\end{enumerate}
The secondary feature in cases ii and iii represents the potential missing feature out of the \CO(J=7--6)/[CI]\CIt\ pair. The sum of the absolute difference between the model and data, hereafter referred to as the residual, is calculated. The model with lowest residual is selected as the best representation of the data, and if either (ii) or (iii) are selected, the primary and secondary features are incorporated into the total model for the full-band spectrum. This process is summarised graphically in Fig.\,\ref{fig:NeutralC_FlowChart}. A key point to successful detection using this regression analysis method is accuracy in the initial parameter estimates input to the fitting routine, and understanding the effect of such initial estimates on the resultant residual. 

\begin{figure}
\centering
\includegraphics{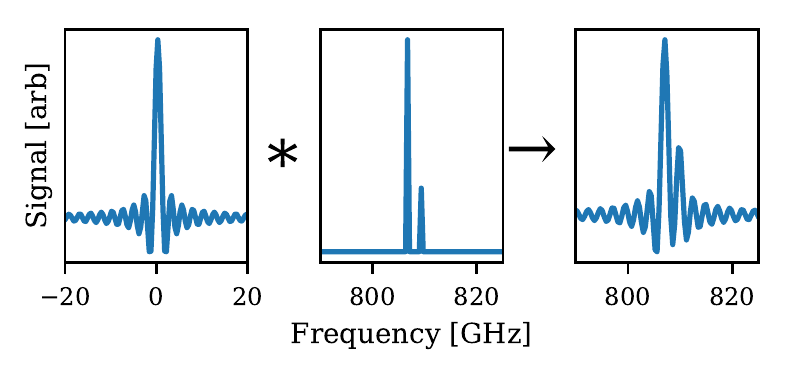}
\vspace{-12pt}
\caption{The \CO(J=7--6)/[CI]\CIt\ profile can be modelled as the convolution of two delta functions, located at their respective rest frequencies, with the instrument line shape.}
\label{fig:convolution}
\end{figure}

The ideal SPIRE FTS instrument line shape (ILS) is a sinc function
\begin{equation}
\centering
\label{eq:sinc}
\mathrm{sinc (\nu)} = \frac{\mathrm{A}\sin((\nu-\mathrm{C})/w)}{(\nu-\mathrm{C})/w}
\end{equation}
where $ \nu $ is frequency, A is the amplitude, C is the frequency centre, and $w$ is a measure of the sinc width equal to the spectral resolution divided by $ \pi $ \citep{spire_handbook}. The typical case of unresolved spectral lines can be modelled as the convolution of the ILS with a delta function located at the emission frequency (Fig.\,\ref{fig:convolution}). This results in a sinc function of fixed width ($w \sim $0.38\,GHz for HR spectra). Thus, for unresolved spectral features, the only parameters that must be determined in the FF spectral line fitting routine are the line centre and the amplitude.

The first parameter of interest is the line centre of the [CI]\CIt\ feature. To predict its position, we assume that both [CI]\CIt\ and \CO(J=7--6) are coupled in the same reference velocity frame. From this assumption, the separation between the \CO(J=7--6) and [CI]\CIt\ features is 2.69\,GHz in the rest frame. For non-relativistic frames with a non-zero radial velocity, v, between source and observer, this separation is modified to
\vspace{-6pt}
\begin{equation}
\label{eq:DeltaNBH}
\centering
\Delta = \left( 1 + \mathrm{v}/c\right)^{-1} \times 2.69 \qquad \mathrm{[GHz]}.\vspace{-6pt}
\end{equation}
This equation has limitations for broadened lines, but is adequate for describing the unresolved spectral features which are the majority of lines in SPIRE spectra. Thus, knowledge of v is required to both locate the \CO(J=7--6) feature and determine the separation of the \CO\ and [CI]\CIt\ features. For this reason only spectra with radial velocity estimates meeting the FF quality threshold are processed using the NCC \citep{FFredshift}. This $ \Delta $ is the value used to position the secondary feature in the NCC algorithm outlined in case ii and iii above.

\begin{figure*}
\centering
\includegraphics{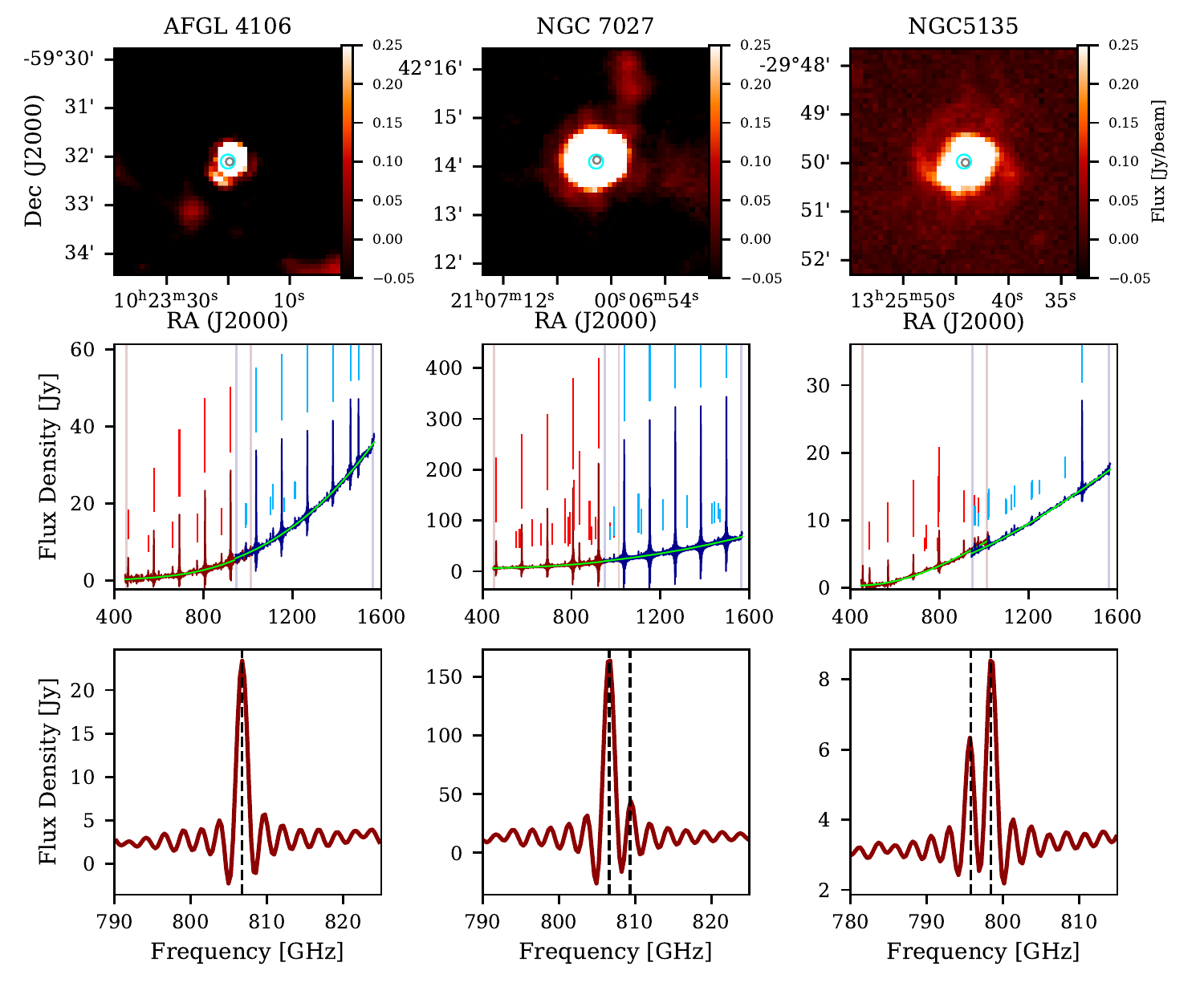}
\vspace{-15pt}
\caption{Example targets of the NCC routine. The top row shows short wavelength (250 $\mu$m) SPIRE photometer maps of carbon rich sources with circles indicating the beams of the central detectors from the SPIRE FTS arrays. The second row shows the corresponding full band SPIRE spectra from the central detectors. The third row is a close up view of the \CO(J=7--6) region with the spectral features present indicated by vertical dashed lines. The first column contains the source AFGL 4106, a post-red supergiant binary, that exhibits no detectable neutral carbon. The young planetary nebula NGC 7027 is shown in the second column and the starburst galaxy NGC 5135 is shown in the third column; these contain faint and strong neutral carbon lines, respectively.}
\label{fig:3cases}
\end{figure*}

\begin{figure}
\centering
\begin{minipage}{1.0\linewidth}
    \includegraphics{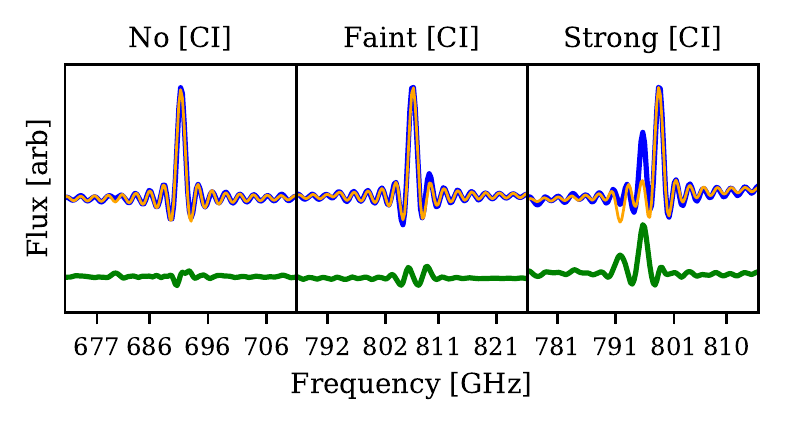}
\vspace{-9pt}
\end{minipage}
\begin{minipage}{1.0\linewidth}
    \includegraphics{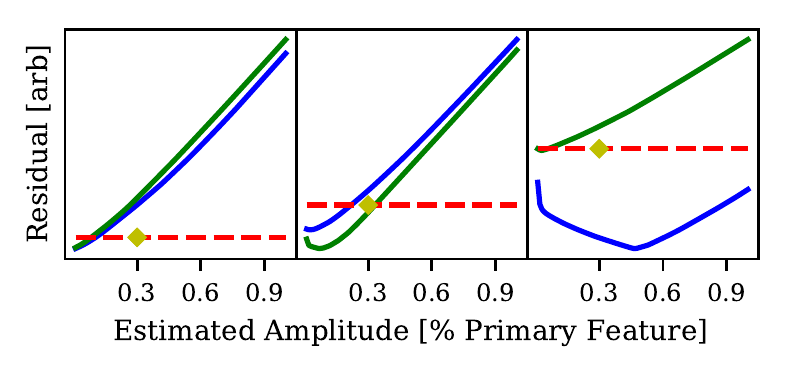}
\end{minipage}
\vspace{-15pt}
\caption{\label{fig:CI_Features}The top row illustrates three test case spectra (dark blue), and resultant spectral fitting for one primary feature (light orange), with the corresponding residual (green). The bottom row shows the resulting residual for different initial amplitude estimates of the secondary feature. Blue and green curves show the resulting residuals obtained when including the secondary sinc feature either to the left (lower frequency, blue) or the right (higher frequency, green) of the primary feature. Red dashed lines show the residual obtained when fitting the primary feature only and are used for reference with the other curves. The diamonds are also shown for reference and illustrate the intersection of the initial 30\% secondary feature amplitude estimate and the single-feature residual. The lowest residual case at the 30\% initial amplitude is taken to be the most likely.}
\end{figure}

\begin{figure}
\centering
\includegraphics{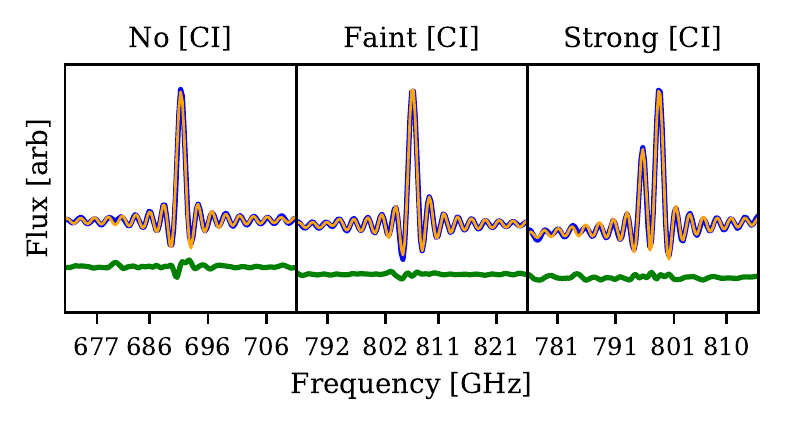}
\vspace{-12pt}
\caption{\label{fig:Final_Fit} Results obtained using the NCC for the no [CI]\CIt\ case (left), weak [CI]\CIt\ case (centre), and strong [CI]\CIt\ case (right).}
\vspace{-9pt}
\end{figure}

The secondary feature amplitude estimate that serves as an input to the minimization routine is the second parameter of interest, and was chosen to be 30\% of the amplitude of the primary feature, based on careful empirical selection and illustrated in Fig.\,\ref{fig:CI_Features}. To see how this parameter was determined, we consider three sources representative of the three scenarios that we wish to accommodate (represented by columns in Figs.\,\ref{fig:3cases}, \ref{fig:CI_Features}, and \ref{fig:Final_Fit}). The first case is a source that contains no detectable neutral carbon feature; a small perturbation on the side lobes gives the impression that an additional feature may be present, but this asymmetry is an artifact within the instrument itself \citep{Naylor:16}. The second case contains a faint [CI] feature adjacent to a prominent \CO(J=7--6) feature, and is representative of the scenario for which the NCC was developed. It should be noted that the NCG 7027 spectrum used for these tests is the standard product available through the HSA and not the Highly Processed Data Product\footnote{\url{http://archives.esac.esa.int/hsa/legacy/HPDP/SPIRE/SPIRE-S/cal_targets/}} (HPDP; \cite{Hopwood15}) variant. The final case contains a strong [CI]\CIt\ feature, and is included for completeness. 

Optimization of the initial amplitude estimate was explored by performing numerous fits to spectral samples from these three cases. The top row of Fig.\,\ref{fig:CI_Features} shows the input spectrum, the model fit, and the associated spectral residual. The model fit is composed of the primary line along with other lines in the spectral sample identified by the main FF routine, and is used to determine line parameters for the primary feature. These parameters were included in a secondary fit where the \CO(J=7--6) and [CI]\CIt\ lines were fit simultaneously. The initial amplitude estimate of the secondary line was varied in terms of percentage of the primary feature amplitude and the resulting output residuals were compared. In the lower row of Fig.\,\ref{fig:CI_Features}, the red dashed lines show the residual of the corresponding example plots above where no secondary feature was fit. These lines are not a function of secondary feature amplitude estimate and are included as a visual reference. The blue and green curves show the residuals obtained by fitting a second sinc 2.69\,GHz (rest frame) to the left (lower frequency), and right (higher frequency), of the primary feature, respectively, while varying the initial amplitude estimate. The diamond on the plot highlights the chosen 30\% secondary feature amplitude estimate. The lowest residual signal at this point indicates the spectral case to which the test spectrum belongs. The left column has the red dashed line with the lowest residual at the horizontal position of the diamond, indicating only the primary feature is the most likely case.  Likewise, the central column has the green curve with the lowest residual at the horizontal position of the diamond, indicating that the secondary feature fit to the right (i.e., [CI]\CIt) is optimal. Finally, in the right column the blue curve is the lowest at the diamond, indicating that the secondary feature fit to the right (i.e., \CO(J=7--6)) is optimal. During this step, the fitting routine tolerances were modified, and the reported residual is for the optimized fit and not the initial guess. If a secondary feature is identified (i.e., case ii or iii provides the lowest residual), the spectral section is fit again with nominal fit tolerance and input parameters from the output of the previous minimization routine. If a fit with the nominal tolerance produces lines within 0.8\,GHz of their expected, velocity corrected frequencies, the lines are added to the main FF model. Performance is dependent on the fitting routine configuration and tolerance, which is specific to observation calibration as a result of the different flux density units (i.e., point source or extended source calibration \citep{spire_handbook}).

In order to test the case where no [CI]\CIt\ is present, we take extra precaution and isolate a \CO(J=6--5) feature, instead of the nominal \CO(J=7--6) feature, with good SNR from the NGC 7027 observation. As demonstrated in Fig.\,\ref{fig:CI_Features}, it can be seen that an initial estimate with an amplitude less than $\sim$10\% of the primary feature always returns a result with a lower residual than fitting the primary feature alone (red dashed line). This result is observed regardless of what side the secondary feature is inserted as would be expected for a feature with a high degree of symmetry. In the second case we isolated a \CO(J=7--6) emission line with a faint [CI]\CIt\ feature. The secondary feature being fit to the right produces a much lower residual than when fit to the left. Once the initial estimate for the amplitude surpasses $\sim$35\%, the 2-sinc fit produces a larger residual than fitting the primary feature alone. For the final case we have isolated a \CO(J=7--6) line with a prominent [CI]\CIt\ feature. In this case the residual is reduced when a second feature is introduced to the right regardless of the initial amplitude estimate.

To detect [CI]\CIt\ while avoiding false positives, the initial amplitude estimate should produce a higher output residual when no [CI]\CIt\ is present and a lower residual when [CI]\CIt\ is present. From the empirical study exemplified in Fig.\,\ref{fig:CI_Features}, it can be concluded that a starting estimate of 30\% primary feature amplitude produces the desired result for these test cases. Fig.\,\ref{fig:Final_Fit} demonstrates the result when the routine outlined is applied to the test case spectral segments.

\begin{figure}
\centering
\includegraphics{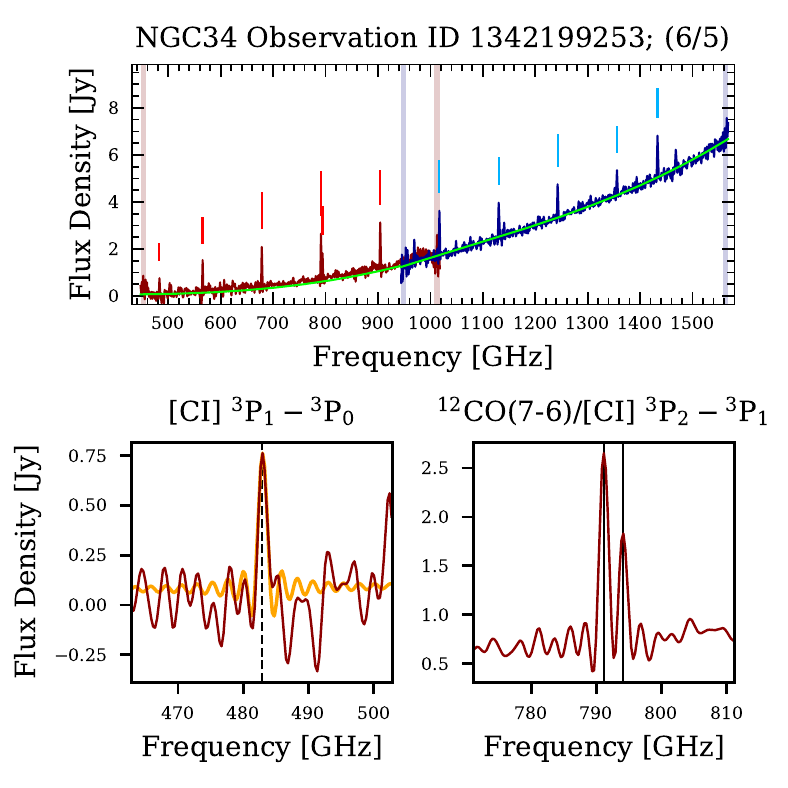}
\vspace{-12pt}
\caption{The top frame shows the FF postcard for an observation of NGC~34 (also known as NGC~17), a galaxy formed from the merger of two galaxies. The bottom-left frame shows a spectral sample (red), the expected frequency of the [CI]\CIo\ feature (dashed line) based on the detected [CI]\CIt\ line frequency, and the resulting sinc line shape fit (orange). The bottom-right frame shows the FF/NCC-detected \CO(J=7--6)/[CI]\CIt\ spectral feature emission frequencies (solid lines) used to determine the expected [CI]\CIo\ feature frequency.\vspace{-12pt}}
\label{fig:CI_predictions}
\end{figure}

Successful detection of the [CI]\CIt\ provides confidence in the potential existence of the associated [CI]\CIo\ feature. The [CI]\CIo\ line is located at a rest frequency of $492.16$\,GHz which puts it at the noisy edge of the SLW band, resulting in unreliable detection by the FF with the NCC turned off. To improve detection of this feature, a dedicated search is performed if the NCC or nominal FF detects a feature residing within 0.8\,GHz of the velocity adjusted rest frequency emission of [CI]\CIt, and no feature is detected by the nominal FF within the same range of the velocity adjusted rest frequency emission of [CI]\CIo. The search process has similarities with the procedure outlined for the [CI]\CIt\ feature. A spectral sample of 40\,GHz is isolated, centered on the expected emission frequency of the [CI]\CIo\ feature, based on the detected frequency of the [CI]\CIt\ feature. An initial amplitude estimate of 30\% that of the [CI]\CIt\ feature is used. If the resulting model produces an SNR $ \ge $5.0 and the center does not deviate by more than 0.8\,GHz from the expected emission frequency, the feature is added to the catalogue (see Fig.\,\ref{fig:CI_predictions}).

The NCC takes place after the main execution of the FF, and operates by modelling an isolated spectral sample as opposed to the full spectral band used in the nominal FF routine. This is a distinct difference worth being aware of since the side-lobes of the ILS produce non-local effects. Thus, strong features outside the spectral sample can influence NCC results. Additionally, the SNR calculation in the NCC uses a window 5\,GHz smaller on either side of the feature than the window used by the nominal FF SNR calculation. This modification is used to accommodate the modelled 40\,GHz spectral sample used by the NCC, in which our testing has shown to produce better results than with a larger sample. With these distinctions in mind, all features added or modified by the NCC are identified in the FF catalogue(s) by the `nccFlag' column.

\section{Validation}
\label{sec:Validation}

Validation of the NCC is accomplished using a combination of manual inspection and redundant detection of neutral carbon features previously detected by the main feature detection routine. This validation analysis is performed exclusively on sparse observations including both point-source and extended-source calibrated spectra, and is used to provide confidence in results obtained from both sparse and mapping observations.

\subsection{Validation of Detected [CI] \texorpdfstring{$\mathrm{^3P_2 - ^3P_1}$}{3P2-3P1} Features}
\label{subsec:CI21}

125 randomly selected point-source calibrated sparse observations with good velocity estimates were manually inspected and evaluated for the presence of [CI]\CIt. These observations were processed by the NCC with any [CI]\CIt\ features detected in earlier steps of the FF masked. The output products were than assessed for valid [CI]\CIt\ detections. The results of this test are shown in Tab.\,\ref{tab:KnownCI}. No false detections occurred, but 4 features failed to be detected resulting in a 97\% detection success rate for the observations tested. The missed features have SNRs ranging from 14.5 to 5.1 with a median value of 6.1, indicating that the missed features were of typically of low SNR. 

\begin{table}
\begingroup
\begin{center}
\newdimen\tblskip \tblskip=5pt
\caption{\label{tab:KnownCI} Statistics from implementing the NCC on 125 spectra where the presence of [CI]\CIt\ is known.}
\nointerlineskip
\small
%
\newdimen\digitwidth
\setbox0=\hbox{\rm 0}
\digitwidth=\wd0
\catcode`*=\active
\def*{\kern\digitwidth}
\newdimen\signwidth
\setbox0=\hbox{+}
\signwidth=\wd 0
\catcode`!=\active
\def!{\kern\signwidth}
%
\tabskip=2em plus 2em minus 2em
\halign to \hsize{\hfil#&**#\hfil& \hfil#\hfil& \hfil#**\hfil&#\hfil\cr
 &\multispan3\hrulefill& \cr
\noalign{\vspace{-8.0pt}}
 &\multispan3\hrulefill& \cr
& & Correct& Incorrect& \cr
\noalign{\vspace{-5.5pt}}
 &\multispan3\hrulefill& \cr
 & CI Present&    *93& 0& \cr
 & No CI Present& *32& 4& \cr
 & Total&         125& 4& \cr
\noalign{\vspace{-7.5pt}}
 &\multispan3\hrulefill& \cr
\noalign{\vspace{-7.5pt}}
}
\end{center}
\endgroup
\vspace{-12pt}
\end{table}

\begin{figure}
\centering
\includegraphics{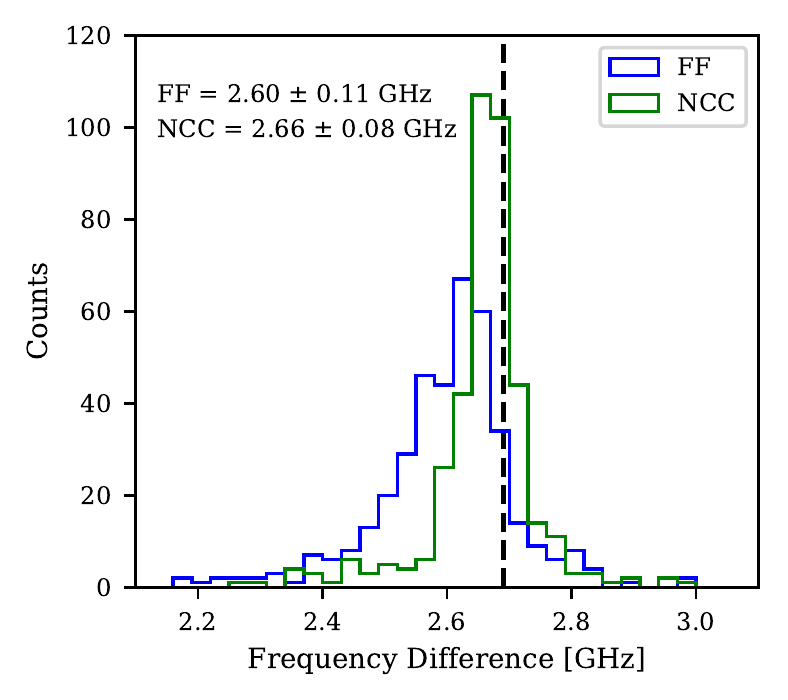}
\vspace{-12pt}
\caption{Velocity adjusted frequency difference between detected \CO(J=7--6) and [CI]\CIt\ lines. The NCC routine (green) produced a distribution of frequency differences closer to the expected 2.69\,GHz than the nominal FF routine (blue).}
\label{fig:FrequencyDifferences}
\end{figure}

As an additional test, features detected in all sparse observation before the execution of the NCC were searched for lines residing within 0.8\,GHz of the velocity adjusted rest frequency of [CI]\CIt. If a feature was found meeting this criteria, the observation was flagged as containing [CI]\CIt. The identified feature was then masked and the observation was processed by the NCC. Of the 439 observations processed in this way, 392 recovered the [CI]\CIt\ feature resulting in an 89\% success rate. These redundant detections by the NCC received associated SNRs which were on average $ \sim $7\% lower than the nominal FF detection. This is explained by the propagation of side-lobes from sinc features outside of the sampled region. Effects caused by features outside of the sampled region are difficult to incorporate into the modelling process and thus result in a higher residual and lower SNR. It was also observed that these redundant detections produced a velocity adjusted frequency difference between the \CO(J=7--6) and [CI]\CIt\ lines that was closer to the expected 2.69\,GHz (Fig.\,\ref{fig:FrequencyDifferences}). Though the NCC produced lower SNRs than the nominal FF routine for these redundant detections, a frequency difference closer to 2.69\,GHz is correlated with higher average SNR for the \CO(J=7--6) and [CI]\CIt\ lines.

Examining the spectra for which [CI]\CIt\ was not recovered reveals the limitations of this routine. Broadly speaking, the most significant limitation is poorly defined local continua. The continuum is strongly influenced by the chosen calibration scheme used to process the spectral data \citep{Wu2013}. In some cases, the poorly defined local continuum is a result of an extended or semi-extended source being processed with the point-source calibration scheme. In cases such as this, the undetected [CI]\CIt\ feature in point-source calibrated spectra can often be found in the complimentary extended-source calibrated spectra. More commonly the continuum is perturbed by noise in the spectrum, while the presence of significant noise features makes line detection even more challenging. A less common limitation is the presence of strong features located just outside of the spectral sample that is under inspection. Specifically, in two observations the side-lobes of strong CH$ ^+ $(J=1--0) emission (835.07\,GHz rest frequency) seem to result in slightly higher residuals when a secondary feature is added to the model of the spectral sample. 

\subsection{Validation of Detected [CI] \texorpdfstring{$\mathrm{^3P_1 - ^3P_0}$}{3P1-3P0} Features}
\label{subsec:CI10}

Only 13 unique [CI]\CIo\ features were detected within sparse observations using the NCC. These features were manually inspected and concluded to be valid detections. The exercise of masking previously detected features and attempting to recover these features using the NCC, as outlined in \S\,\ref{subsec:CI21}, was repeated for [CI]\CIo\ lines. Of the 321 observations processed in this way, the NCC successfully recovered 318 of the masked [CI]\CIo\ features. The three exceptions were due to the absence of a [CI]\CIt\ detection and thus the NCC did not attempt to find the masked feature.

\section{Results}
\label{sec:Results}
\begin{figure}
\centering
\begin{minipage}{1.0\linewidth}
\centering
\includegraphics{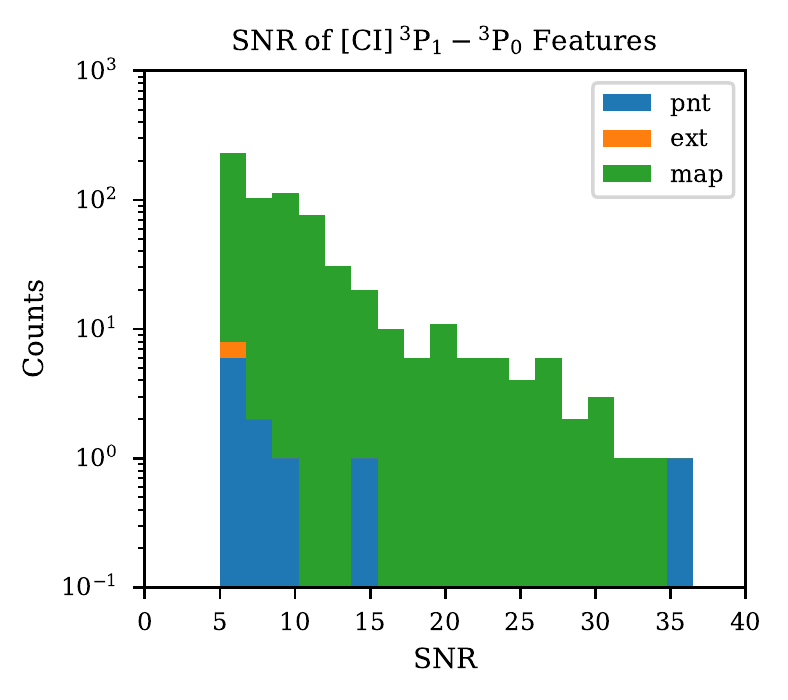}
\includegraphics{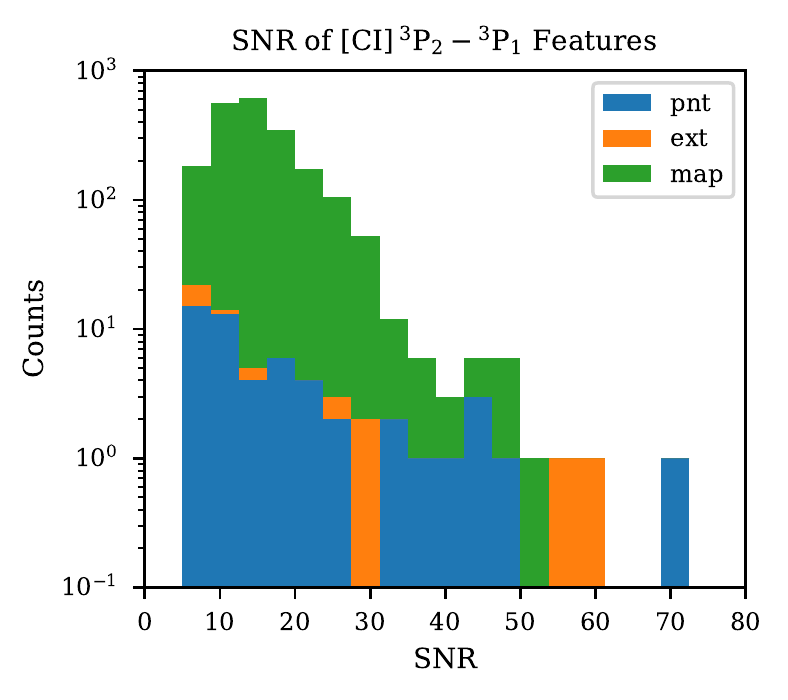}
\includegraphics{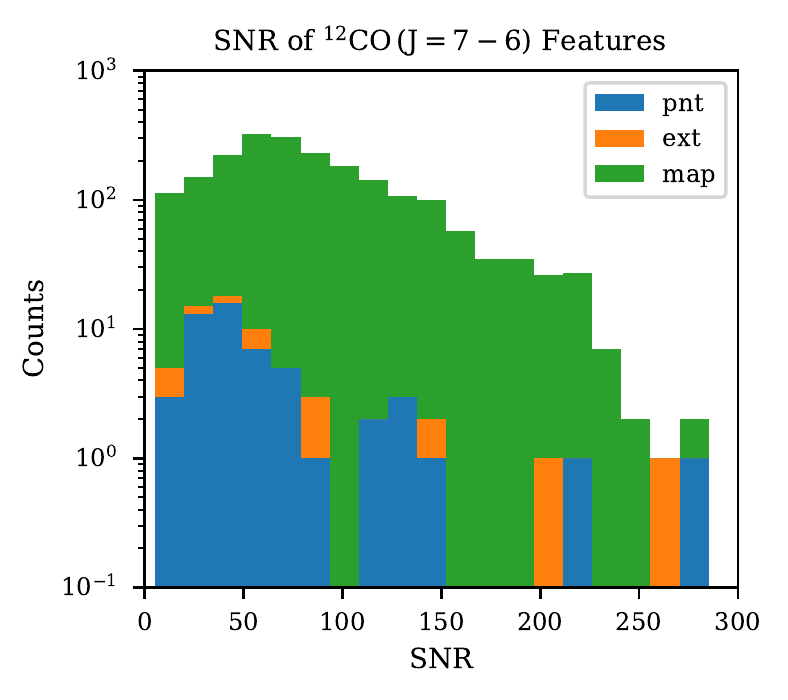}
\end{minipage}
\caption{The SNR distribution of features detected or modified by the NCC. Features contained within point-source calibrated and extended-source calibrated sparse observations are indicated with blue and green bars, respectively. Features contained within mapping observations are indicated with red bars.\vspace{-12pt}}
\label{fig:NCC_SNR}
\end{figure}

The NCC was applied to all HR SPIRE observations processed by the FF (see \citetalias{FFtech}). The number of features detected or modified/confirmed by the NCC for mapping observations and the central detectors of sparse observations are summarised in Tab.\,\ref{tab:NCC_Results}. The SNR of these features is presented in Fig.\,\ref{fig:NCC_SNR}. Recall that for [CI]\CIt\ the NCC simultaneously fits both the [CI]\CIt\ and the rotational $ ^{12} $CO(J=7--6) feature. Thus, for every detected [CI]\CIt\ feature, there is a corresponding $ ^{12} $CO(J=7--6) feature confirmation, and vice versa. The extension to off-axis detectors of sparse observations is presented in a companion paper \citep[\citetalias{FFlineID}:][]{FFlineID}.

As would be expected, the SNRs of \CO(J=7--6) features are generally higher than the SNRs of [CI]\CIt\ features. Similarly, the SNRs of [CI]\CIt\ features are typically higher than [CI]\CIo\ features. 

Significantly more [CI] features were detected in the mapping observation dataset than the sparse datasets. This is not unexpected by virtue of the fact that convolved spectral maps contain on the order of a few hundred pixels per observation, each containing a spectrum processed by the FF.   

\begin{table}
\begingroup
\begin{center}
\newdimen\tblskip \tblskip=5pt
\caption{\label{tab:NCC_Results} The number of spectral features detected or modified/confirmed by the NCC for mapping observations and the central detectors of sparse observations. The columns `point cal' and `extended cal' refer to features contained within the point-source calibrated and extended-source calibrated spectra of sparse observations, respectively, while the `mapping' column refers to features contained within mapping observations.}
\nointerlineskip
\small
%
\newdimen\digitwidth
\setbox0=\hbox{\rm 0}
\digitwidth=\wd0
\catcode`*=\active
\def*{\kern\digitwidth}
\newdimen\signwidth
\setbox0=\hbox{+}
\signwidth=\wd 0
\catcode`!=\active
\def!{\kern\signwidth}
%
\tabskip=2em plus 2em minus 2em
\halign to \hsize{\hfil#&\hfil*#\hfil& \hfil#\hfil& \hfil#\hfil& \hfil#*\hfil&#\hfil\cr
 &\multispan4\hrulefill& \cr
\noalign{\vspace{-8.0pt}}
 &\multispan4\hrulefill& \cr

 &  & point cal& extended cal& mapping \cr 
\noalign{\vspace{-5.5pt}}
 &\multispan4\hrulefill& \cr
\noalign{\vspace{3pt}}
 & [CI]*\CIo\ & 11& *2& *\,620& \cr 
\noalign{\vspace{3pt}}
 & [CI]*\CIt\ & 53& 14& 2\,007& \cr 
\noalign{\vspace{3pt}}
 & $^{12}$CO*(J=7--6) & 53& 14& 2\,007& \cr  
\noalign{\vspace{3pt}}
\noalign{\vspace{-7.5pt}}
 &\multispan4\hrulefill& \cr
\noalign{\vspace{-7.5pt}}
}
\end{center}
\endgroup
\vspace{-12pt}
\end{table}

\section{Conclusions}
\label{sec:Conclusion}
Automated detection of the [CI]\CIt\ emission line is challenging due to the limitations imposed by the spectral resolution of the SPIRE FTS and the close spectral proximity of the \CO(J=7--6)/[CI]\CIt\ pair. Using a preliminary radial velocity estimate, it is possible to predict when the NCC algorithm should be applied, and what the line centre estimate should be for the subsequent fitting procedure. The carefully selected initial amplitude estimate of secondary features within the NCC routine helps to detect the presence of [CI]\CIt\ emission while at the same time limiting false [CI]\CIt\ detection in the automated SPIRE spectral FF line fitting routine. The NCC has demonstrated a $ \ge $89\% successful detection rate of [CI]\CIt\ emission blended with a rotational \CO(J=7--6) feature, and potentially greater success can be achieved with improved continuum parameterization. 

Automated [CI]\CIo\ detection can also be challenging resulting from a typically low amplitude feature embedded in the noisy low frequency edge of the SLW band. The NCC finds a natural extension towards detecting [CI]\CIo\ features with successful [CI]\CIt\ detection. Tests using redundant detection demonstrate a near 100\% detection rate which seems to only be limited by simultaneous detection of the [CI]\CIt\ feature. No false detections have been observed for either [CI] feature when using the NCC. 

Features detected or modified by the NCC have been incorporated into public FF products and can be identified using the `nccFlag' column in the feature catalogues. We stress that the NCC is conducted after the main feature detection routine and operates by modelling a spectral sample which cannot account for the effects of strong emission features outside of, but in close spectral proximity to, the spectral sample region. Though detection is considered reliable, the FF and NCC are not intended to be a replacement for detailed spectral analysis.  

\section*{Acknowledgements}
\emph{Herschel} is an ESA space observatory with science instruments provided by European-led Principal Investigator consortia and with important participation from NASA\@. This research acknowledges support from ESA, CSA, CRC, CMC, and NSERC\@. 

This research has made use of the SciPy (\url{www.scipy.org}) and Astropy (\url{www.astropy.org}) Python packages. Table formatting in this paper followed the {\it Planck} Style Guide \citep{PlanckStyle}.
\nocite{2020SciPy-NMeth, astropy:2013, astropy:2018}\\

\section*{Data Availability}
The \textit{Herschel} SPIRE Spectral Feature Catalogue has been assigned an ESA Digital Object Identifier (DOI) and is available at: \href{https://doi.org/10.5270/esa-lysf2yi}{doi.org/10.5270/esa-lysf2yi}. The FF code and all FF products are publicly available via the \textit{Herschel} Science Archive. 




\bibliographystyle{mnras}
\bibliography{mnras_NeutralCarbon} 







\bsp	
\label{lastpage}
\end{document}